\documentclass[aps,prb,reprint,superscriptaddress]{revtex4-1}

\usepackage{mhchem} 
\usepackage{graphicx}
\usepackage{hyperref}
\newcommand{\tc}{Wm$^{-1}$K$^{-1}$ }

\begin{document}

\title{Phonon anharmonicity, lifetimes and thermal transport in CH$_3$NH$_3$PbI$_3$ from many-body perturbation theory}



\author{Lucy D. Whalley}
\affiliation{Department of Materials, Imperial College London, Exhibition Road, London SW7 2AZ, United Kingdom}

\author{Jonathan M. Skelton}
\affiliation{Department of Chemistry, University of Bath, Claverton Down, Bath BA2 7AY, United Kingdom}

\author{Jarvist M. Frost}
\affiliation{Department of Materials, Imperial College London, Exhibition Road, London SW7 2AZ, United Kingdom}
\affiliation{Department of Chemistry, University of Bath, Claverton Down, Bath BA2 7AY, United Kingdom}

\author{Aron Walsh}
\email[]{a.walsh@imperial.ac.uk}
\affiliation{Department of Materials, Imperial College London, Exhibition Road, London SW7 2AZ, United Kingdom}
\affiliation{Department of Chemistry, University of Bath, Claverton Down, Bath BA2 7AY, United Kingdom}
\affiliation{Global E$^3$ Institute and Department of Materials Science and Engineering, Yonsei University, Seoul 120-749, Korea}


\date{\today}

\begin{abstract}
Lattice vibrations in CH$_3$NH$_3$PbI$_3$ are strongly interacting, with double-well instabilities present at the Brillouin zone boundary. Analysis within a first-principles lattice-dynamics framework reveals anharmonic potentials with short phonon quasi-particle lifetimes and mean-free paths. The phonon behaviour is distinct from the inorganic semiconductors GaAs and CdTe where three-phonon interaction strengths are three orders of magnitude smaller. The implications for the applications of hybrid halide perovskites arising from thermal conductivity, band-gap deformation, and charge-carrier scattering through electron-phonon coupling, are presented. 
\end{abstract}


\maketitle

Hybrid halide perovskites have been the subject of intensive investigation 
due to their strong photovoltaic action.\cite{Stranks2015b}
While solar-cell device efficiencies are high,
our understanding of the materials properties remains limited in comparison.
Here we address the anharmonic nature of phonons in \ce{CH3NH3PbI3}
from a theoretical perspective, with a particular focus on 
phonon interactions, lifetimes,
and coupling to the electronic structure.

\textit{Lattice vibrations of hybrid perovskites.}
Vibrational spectroscopy is a valuable tool in materials characterisation. 
As such there have been multiple reports concerning the 
IR and Raman activity of \ce{CH3NH3PbI3}.\cite{Quarti2013,Ledinsky2015,Bakulin2015a,PerezOsorio2015a,Glaser2015a,Brivio2015a}
Chemical breakdown to \ce{PbI2} is a concern, but once careful measurements are made,
satisfactory agreement can be obtained between 
first-principles theory and experiment.\cite{Brivio2015a,PerezOsorio2015a}
As expected from the large difference in atomic mass, vibrations of the \ce{PbI3-} framework are found at
lower energy (0 -- 5 THz)
with \ce{CH3NH3+} vibrations at higher energy (8 -- 100 THz); 
however, significant coupling is found between the two,
 as explored in our earlier work\cite{Brivio2015a}
and observed in neutron scattering measurements on \ce{CH3NH3PbBr3}.\cite{Swainson2015}

\begin{figure}[h!]
\includegraphics[width=8.6cm]{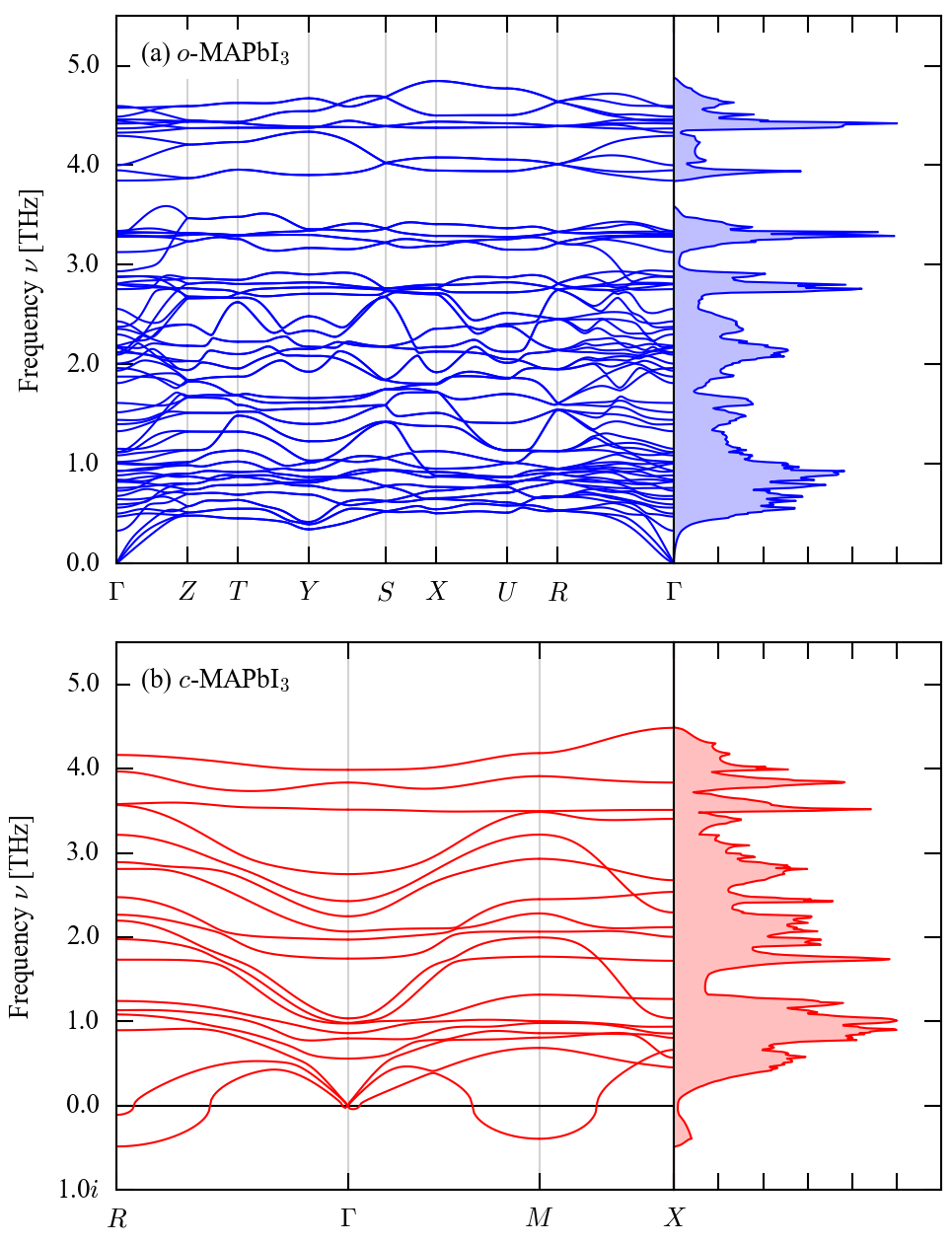} 
\caption{
(Color online) 
Harmonic phonon dispersion up to 5 THz in the (a) low-temperature (orthorhombic) and (b) high-temperature (cubic) 
perovskite phases of \ce{CH3NH3PbI3}. 
They are calculated from lattice dynamics using forces from density functional theory (PBEsol
electron exchange and correlation).
}
\label{fig-phonon}
\end{figure}

\textit{Harmonic phonon dispersion.}
The potential energy ($U$) of a crystal can be expanded as a Taylor series of ionic displacements ($r$).
The static crystal potential ($U_0$) has no bearing on the dynamics, and for 
a relaxed structure the $\frac{d U}{d r}$ term is zero.
In the harmonic approximation only the $\frac{d^2 U}{d r^2}$ term is considered. 
A dynamical matrix so constructed is positive definite for an equilibrium
structure, and the phonon eigenmodes from its solution are orthogonal (and therefore
non-interacting). 
While a number of salient features of the thermal physics can be reproduced
by harmonic lattice dynamics, the resulting phonon modes
have  temperature independent frequencies and possess infinite lifetimes.
The quasi-harmonic approximation (QHA), harmonic potentials  are calculated over a specified volume range, was developed to deal with the first issue,
whilst phonon-phonon interactions must be considered for
an explicit treatment of the second issue. 

The harmonic phonon dispersion for \ce{CH3NH3PbI3}
in the low-temperature orthorhombic and high-temperature cubic phases is shown in Figure \ref{fig-phonon}.
We use the same computational setup previously reported\cite{Brivio2015a}
based on \textsc{Phonopy}\cite{Togo2015a}, \textsc{VASP}\cite{Kresse1996a}
and the PBEsol exchange-correlation functional\cite{Perdew2008a} (see Supplemental Information\cite{si}).
There is significant dispersion across the vibrational Brillouin zone
in the low frequency modes.
We focus on the region up to 5 THz where vibrations of the \ce{PbI3^-} framework
are found.
While IR and Raman spectroscopy probe the $\Gamma$-point ($q$ = 0) modes, this
represents a small fraction of the possible lattice vibrations.   
All of the phonon modes in the orthorhombic phase have zero or positive frequencies. 
Two imaginary frequency acoustic modes are found in the cubic phase, centred around the 
\textit{R} ($q = \frac{1}{2},\frac{1}{2},\frac{1}{2}$) and \textit{M} ($q
= \frac{1}{2},\frac{1}{2},0$) special points.  
These indicate the presence of a saddle-point in the potential-energy surface,
and thus that the structure is not dynamically stable.  
Such ``soft" or ``imaginary" modes have been recently observed in inelastic X-ray scattering
measurements of the phonon dispersion.\cite{Beecher2016,Sarg2016}

\begin{figure}
\includegraphics[width=8.0cm]{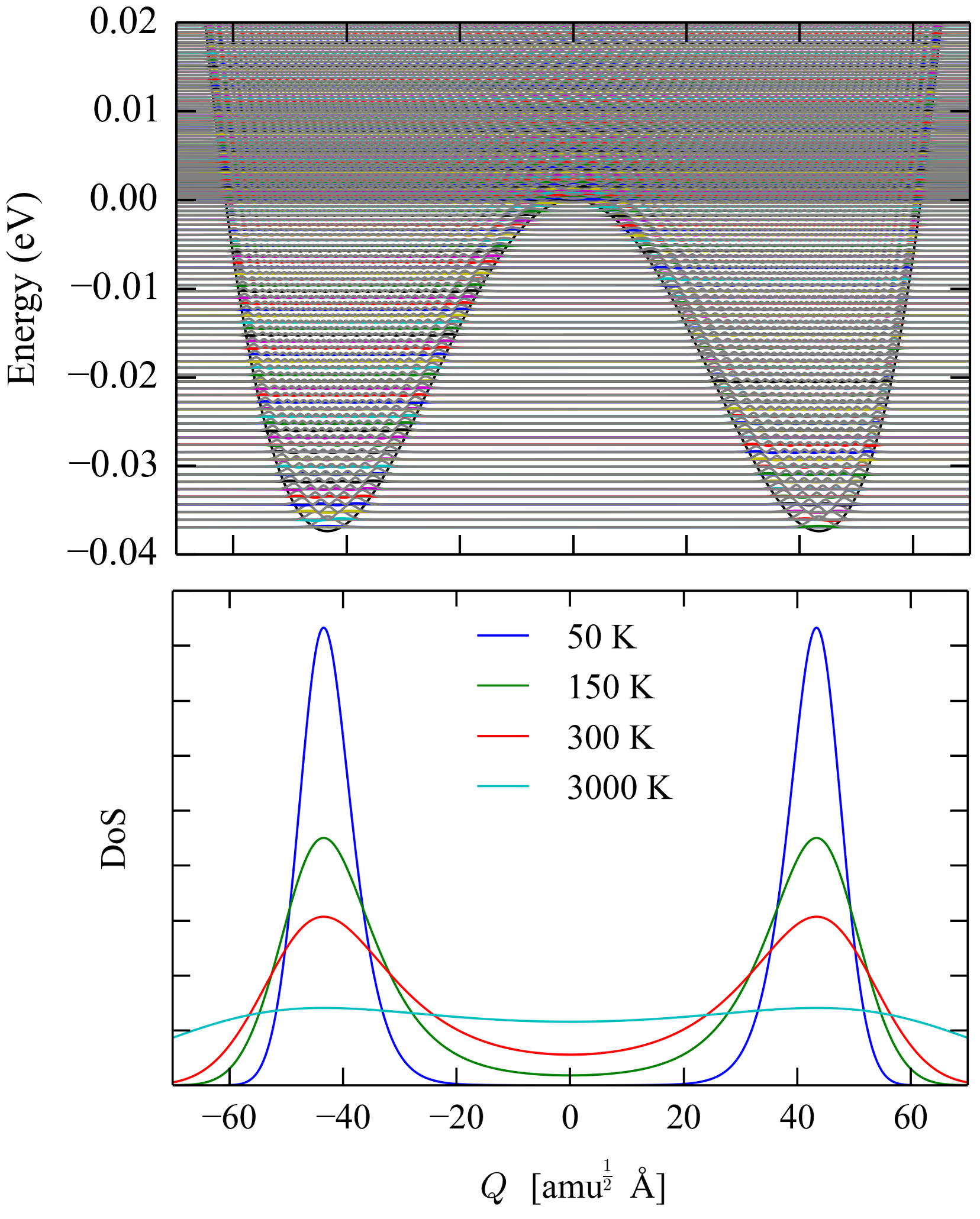}
\caption{
(Color online) 
Potential energy surface from frozen-phonon calculations of the imaginary eigenmode 
present at the \textit{R} point in the high-temperature perovskite phase of  
 \ce{CH3NH3PbI3}.  
$Q$ represents the normal mode coordinate (phonon amplitude).  
The solution of a 1D Schr\"{o}dinger equation (states shown as horizontal lines with wavefunctions
in the upper panel) are used to generate a thermalised probability density of states (DoS; lower panel). 
}
\label{fig-softy}
\end{figure}

\textit{Soft phonon modes.}
The imaginary acoustic phonon modes at \textit{R} and \textit{M}
are zone-boundary instabilities characteristic of the perovskite crystal structure.\cite{Woodward1997,Howard2006} 
They are associated with collective tilting of the corner-sharing octahedral
framework, as observed in molecular dynamics.\cite{Frost2014b,Quarti2015a} 
The same instabilities have been reported in CsSnI$_3$,\cite{Huang2014b,Patrick2015a} 
where they persist even in QHA calculations.\cite{Silva2015}
These zone-boundary motions can be described within a computationally-tractable
$2 \times 2 \times 2$ supercell expansion of the cubic perovskite lattice. 
Within the frozen-phonon approximation, we map out the potential energy surface for 
displacement along the imaginary zone-boundary eigenvectors
(Figure \ref{fig-softy}). 
By following the imaginary mode eigenvectors to map the potential-energy
surface, we are assuming that both the mode
eigenvector and potential-energy surface are set by the crystal symmetry. 
The result is a characteristic double-well potential, where the cubic perovskite
structure is a saddle point between two equivalent broken-symmetry solutions. 
The barriers are significant (37 and 19 meV for the \textit{R} and \textit{M} modes, respectively) and is comparable to $k_{B}T$, so 
order-disorder behaviour is expected.\cite{Dove1997}

The phase transitions from cubic to tetragonal and orthorhombic perovskite structures 
can be understood as a condensation 
of the \textit{R} and subsequently \textit{M} modes.\cite{Howard2006}
Similar transitions are observed in \ce{CsPbCl3}.\cite{Fujii1974}
To further understand the physical behaviour we have solved 
the time-independent Schr\"{o}dinger equation describing the nuclear motion in
this 1D double-well potential.
The procedure, outlined in Ref. \onlinecite{Skelton2016}, 
makes a single-phonon approximation and neglects coupling to 
other phonon modes. 
It assumes that energetic cross-terms from interaction with the other
 modes are small, which is expected for such rigid octahedral tilts.
A similar scheme, based on a similar independent-mode approximation, has recently been reported
by Adams and Passerone.\cite{adams2016insight}

The resulting eigenstates 
form a partition function which can be associated
with a renormalised harmonic frequency that 
reproduces the thermodynamic contribution of the anharmonic
system.\cite{Skelton2016}
Values of 0.08 THz (\textit{M}) and 0.10 THz (\textit{R}) are found
for $T$ = 300 K.
Since the phonon occupation is governed by 
Bose-Einstein statistics, these low energy modes
are highly populated, as has been evidenced in
X-ray scattering experiments.\cite{Beecher2016,Sarg2016}

\begin{figure*}
\includegraphics[width=17.2cm]{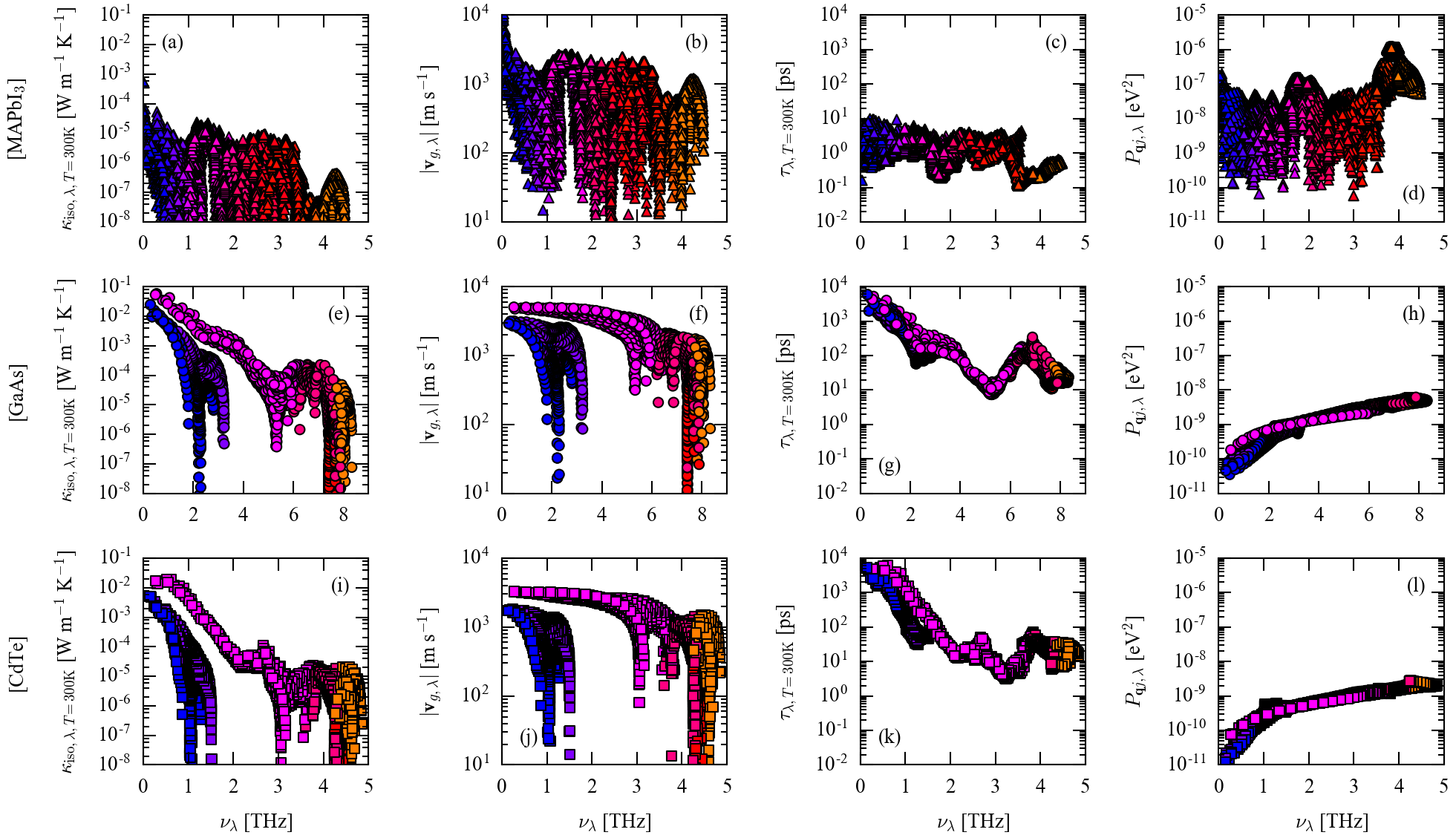} 
\caption{
(Color online) 
Results from anharmonic lattice dynamics calculations on \ce{CH3NH3PbI3}, GaAs and CdTe.
Average phonon modal interaction strength ($P$) and group velocity (\textbf{v}), 
as well as the T = 300 K values of modal lifetime ($\tau$) and thermal conductivity ($\kappa$).
Only the 0 -- 5 THz range is shown for \ce{CH3NH3PbI3}.
The data points are colored according to the band index.
}
\label{fig-breakdown}
\end{figure*}

\textit{Phonon lifetimes and mean-free paths.}
We next consider three-phonon interactions via a peturbative many-body expansion as implemented in \textsc{Phono3py}.\cite{Togo2015,Skelton2014,Togo2015a}
To provide a reference point, we have performed equivalent calculations on the inorganic semiconductors GaAs and CdTe
(both zincblende-type structures), with the results
compared in Figure \ref{fig-breakdown}.
The difference in behaviour is striking with the average strength 
of three-phonon interactions three orders of magnitude larger in \ce{CH3NH3PbI3},
which results in lifetimes ($\tau$) three orders of magnitude \textit{shorter} in the hybrid perovskite. 
While phonon mean free paths
of up to 10 $\mu$m are found in CdTe and GaAs, 
for \ce{CH3NH3PbI3} the limit is 10 nm. 

\begin{figure}
\includegraphics[width=7.5cm]{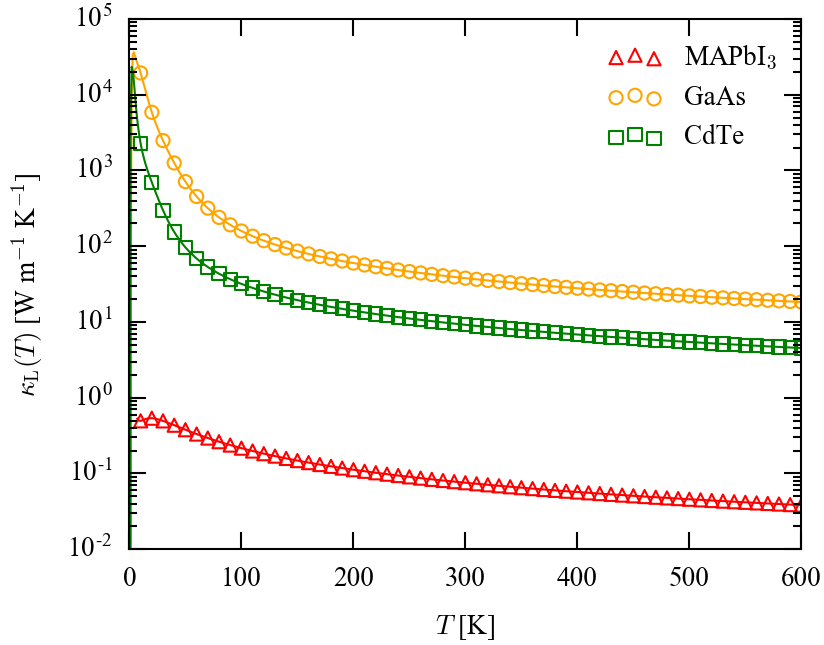} 
\caption{
(Color online) 
Lattice thermal conductivity of \ce{CH3NH3PbI3}, GaAs and CdTe
calculated from three-phonon interactions within the relaxation time approximation (excluding isotope effects).
The calculated values of 38 \tc (GaAs) and 9 \tc (CdTe) at $T$ = 300 K compare well to the measured
values of 45 \tc and 7 \tc, respectively.\cite{Madelung2003}
The corresponding value calculated for  \ce{CH3NH3PbI3} is 0.05 \tc.
}
\label{fig-thermal}
\end{figure}

\textit{Thermal conductivity.}
Lattice thermal conductivity can be expressed as the tensor product of the modal heat capacity ($C_V$), 
group velocity (\textbf{v}) and  phonon mean free path ($\Lambda = \textbf{v} \tau$)
summed over all modes ($\lambda$) and averaged over wavevectors ($q$).
%
%
The result, based upon the values determined from the anharmonic 
lattice-dynamics calculations,
is that \ce{CH3NH3PbI3} is a thermal insulator in comparison to GaAs and CdTe.
The combination of short mode lifetimes and low group velocities (Figure \ref{fig-breakdown}) results in a low averaged thermal conductivity
to 0.05 \tc at $T$ = 300 K (Figure \ref{fig-thermal}).
The `ultra-low' thermal conductivity is in agreement with previous calculations 
and experiments that highlighted potential applications for heat-to-electricity 
conversion in thermoelectric devices.\cite{Pisoni2014,He2014,Mettan2015}

Low-lying acoustic modes are responsible for conducting 
the majority of the heat in CdTe and GaAs.  
Due to the 
unusually short lifetimes of these modes in \ce{CH3NH3PbI3}, it is the cage modes 
in the 1--3 THz window that make the dominant contribution.
While the molecular vibrations (up to 100 THz) influence the lifetimes and mean-free paths of the low-frequency inorganic cage modes, they do  
not directly contribute to thermal transport.

\textit{Carrier scattering.}
The scattering of charge carriers (electrons and holes) in semiconductors is determined by the sum of the rates 
of all possible processes.
%
It is common to identify the dominant scattering mechanism in a sample by the temperature dependence of the transport properties.
While simple textbook relationships exist, 
they are usually for idealised systems, e.g. for harmonic vibrations and parabolic electronic bands.
Hybrid perovskites are non-standard semiconductors: they are mechanically soft, the vibrations are anharmonic, the bands are non-parabolic,\cite{Brivio2014a}
and the dielectric constants are strongly temperature dependent.\cite{Onoda-yamamuro1992}
Deviations from basic models should be expected.

Analysis of photo-conductivity data pointed towards dominant acoustic deformation potential scattering in hybrid perovskites
due to a $T^{-\frac{3}{2}}$ dependence of the scattering time for 150 $<$ $T$ $<$ 300 K.\cite{Karakus2015}
Similar results have been reported from Hall measurements.\cite{Yi2016a}
As the lattice volume fluctuates 
in thermal equilibrium so do the valence and conduction band energies.\cite{Bardeen1950a}
In support of this model, the band gap deformation potential has been calculated to be large\cite{Brivio2014}
and the elastic constants are small.\cite{Sun2015f}
However, for heteropolar materials optical phonon scattering is usually the
dominant mechanism at high temperatures.\cite{Cardona1987} 
As hybrid perovskites possess low-frequency optical phonons, 
the cross-over temperature
is likely to be below 300 K.\cite{Leguy2016}
Recent analysis of emission line broadening highlighted the role of longitudinal optical phonons (the Fr\"{o}hlich interaction), 
where the photoluminescence linewidth at $T$ = 300 K was broken down into inhomogeneous broadening (26 meV) 
and optic-mode broadening (40 meV) components.\cite{Wright2016}
Numerical simulations of scattering kinetics clearly show that the $T^{-\frac{3}{2}}$ behaviour 
can be explained by optic mode events.\cite{Filippetti2016} 

\textit{Anharmonic band gap deformation. }
A common method used to calculate electron-phonon coupling is density-functional perturbation theory (DFPT).\cite{Baroni2001}
DFPT assumes and requires a small harmonic response for the perturbative treatment to be correct.
%
For anharmonic phonon modes, where the range of motion is large, these assumptions do not
hold. 
Previous studies on hybrid perovskites have only considered 
(positive-frequency) harmonic phonons.\cite{Kawai2015,Wright2016}

We return to earlier theories of electron-phonon interaction based on the
phonon-frozen approximation and adiabatic decoupling of the nuclear and electronic
degrees of freedom.\cite{sham1963electron,monserrat2013anharmonic}
To estimate the effect of the anharmonic potential-energy surface on 
the electronic structure of \ce{CH3NH3PbI3}, 
we calculated the change in 
band gap $\Delta E_g (Q)$ with respect to the imaginary-mode phonon amplitude.
A similar band gap deformation is found for the \textit{M} and \textit{R} modes. 
The effect of tilt angle on band gap has been explored in other
studies;\cite{Filip2014a,Amat2014a} however, the present analysis quantifies it
for a collective phonon mode.
We find that $E_g(Q)$ is well described as quadratic over small $Q$, but
required a biquadratic term to reproduce the correct behaviour at large $Q$. 


The expectation value of $E_g$ as a function of temperature ($T$) can be written as
\begin{equation}
    E_g(T) = \left< \chi(Q,T) \,\rvert\, E_g(Q) \,\lvert\, \chi(Q,T) \right>
\end{equation}
where $\chi$ is the thermally-populated vibrational wavefunction 
obtained from solving
the 1D Schr\"{o}dinger equation (Figure \ref{fig-softy}).
All calculations are reduced to the collective phonon
coordinate $Q$, making them computationally tractable.
This, naturally, means that all cross terms are discounted. 

Summing $\Delta E_g (Q)$ multiplied by the vibrational probability density
along $Q$ yields a thermally-averaged electron-phonon coupling for each mode. 
A biquadratic fit to the deformation potential was found to be essential,
due to the considerable contribution from the large $Q$ component of the
wavefunction; using a harmonic approximation for $\Delta E_g (Q)$ led to solutions twice as
large. 
This procedure does not assume that the wavefuntion is centered around $Q=0$.
%
We estimate a positive band gap shift of 35.5 meV (\textit{R} mode) and 27.9
meV (\textit{M} mode) at $T$ = 300 K, which is comparable in magnitude to the measured
broadening of 40 meV.\cite{Wright2016}
The anharmonic electron-phonon coupling of the soft mode in halide perovskites
is therefore considerable, which merits further investigation.

In summary, we have explored the anharmonic nature of
the phonons in \ce{CH3NH3PbI3} and their effect on the physical properties
of the material. 
We have predicted the existence of double-well potentials associated with octahedral tilting, 
provided insights into the strength of the phonon-phonon interactions and thermal transport, and 
highlighted the role of anharmonicity in electron-phonon interactions. 
We did not discuss the rotational activity of \ce{CH3NH3+},
which has been the subject of thorough investigation,\cite{Swainson2015,Leguy2015b,Chen2015s,Bakulin2015a}
and can be considered as an additional anharmonic perturbation. 
Neither have we discussed ion migration, which is a process that can also contribute to thermal properties including 
non-equilibrium thermoelectric power.\cite{Allnatt1961}
In conventional photovoltaic materials such as GaAs 
both the electronic and phonon mean-free paths can exceed 1 $\mu$m,\cite{Cardona1987}
while for \ce{CH3NH3PbI3} the path (limited to 10 nm) is much shorter,
which may have important implications for hot-carrier cooling
and non-radiative recombination processes during solar-cell operation. 

\textbf{Data Access Statement} 
The crystal structures and phonon data are available at \url{https://github.com/WMD-group/Phonons}. 
Codes to estimate anharmonic electron-phonon coupling are available from
\url{https://github.com/jarvist/Julia-SoftModeTISH-DeformationPotential}. 

\begin{acknowledgments}
We thank F. Brivio for preliminary phonon computations,
and J. Buckeridge for assistance with the soft-mode analysis. 
This work was funded by the EPSRC (grant nos. EP/M009580/1, EP/K016288/1,	EP/L01551X/1 and EP/K004956/1), 
the Royal Society and the ERC (grant no. 277757).
Calculations were performed on the UK Archer HPC facility, accessed through membership of the UK HPC Materials Chemistry Consortium (EPSRC grant no. EP/L000202) and the SiSu supercomputer at the IT Center for Science (CSC), Finland, via the Partnership for Advanced Computing in Europe (PRACE) project no. 13DECI0317/IsoSwitch. We also made use of the Balena HPC facility at the University of Bath, which is maintained by Bath University Computing Services.
\end{acknowledgments}


%

\end{document}